\documentclass[12pt,preprint]{aastex}

\begin{document}

\title{The Peculiar Chemical Inventory of NGC~2419 --  An Extreme 
Outer Halo ``Globular Cluster''
\altaffilmark{1}}

\author{Judith G. Cohen\altaffilmark{2},
Wenjin Huang\altaffilmark{3} and Evan N. Kirby\altaffilmark{2,4}  }

\altaffiltext{1}{Based in part on observations obtained at the
W.M. Keck Observatory, which is operated jointly by the California
Institute of Technology, the University of California, and the
National Aeronautics and Space Administration.}

\altaffiltext{2}{Palomar Observatory, Mail Stop 249-17,
California Institute of Technology, Pasadena, Ca., 91125,
jlc(enk)@astro.caltech.edu}

\altaffiltext{3}{Brion Technologies Inc., 4211 Burton Drive, Santa
Clara, Ca. 95054, wenjin.huang@brion.com}

\altaffiltext{4}{Hubble Fellow}

\begin{abstract}

NGC~2419 is a massive outer halo Galactic globular cluster whose stars
have previously been shown to have somewhat peculiar abundance
patterns.  We have observed seven luminous giants that are members of
NGC~2419 with
Keck/HIRES at reasonable SNR.  One of these giants 
is very peculiar,
with an extremely low [Mg/Fe] and high [K/Fe] but normal abundances of
most other elements.  The abundance pattern does not match the
nucleosynthetic yields of any supernova model.  The other six stars
show abundance ratios typical of inner halo Galactic globular
clusters, represented here by a sample of giants in
the nearby globular cluster M30.  
Although our measurements show that NGC~2419 is unusual in
some respects, its bulk properties do not provide compelling evidence
for a difference between inner and outer halo globular clusters.

\end{abstract}


\section{Introduction \label{section_intro} }

NGC~2419 is a globular cluster (GC) in the outer halo of the Galaxy at
a distance of 84~kpc\footnote{The distance and other parameters for NGC~2419
  are from the 2010 version of the on-line GC database of
  \cite{harris96}, based on the photometry of \cite{harris97}.}.
\cite{sigmav09} determined a velocity dispersion for this cluster
based on 40 stars which leads to $M/L = 2.05\pm0.50~M_{\odot}/L_{\odot}$, a normal
value for an old stellar system, with no evidence for  dark
matter at the present time.  \citet*{cls10} further asserted that this cluster could not
have formed in a now-defunct dark matter halo.  \cite{harris97}
obtained a deep CMD of NGC~2419 with HST/WFPC2 and demonstrated that
there is no detectable difference in age between it and M92, an
ancient, well-studied inner halo GC of comparable metallicity.
Furthermore, there is no evidence from CMDs of multiple stellar
populations in NGC~2419.

However, some of the characteristics of NGC~2419 are anomalous for a
GC.  It has the highest luminosity ($M_V \sim -9.6$~mag) of any GC
with Galactocentric radius $R > 20$~kpc, higher than all other GCs
with $R > 15$~kpc with the exception of M54, which is believed to represent
the nuclear region of the Sgr dSph galaxy now being accreted by the Milky Way
\citep{ibata95,sarajedini95}.  NGC~2419 also has an unusually large
half-light radius ($r_h = 19$~pc) and core radius ($r_c = 7$~pc)
for a massive Galactic GC and a relaxation time which exceeds the Hubble
time, also unusual for a GC.  Every other luminous GC in the outer
halo is considerably more compact.  The 2010 version of the on-line
database of \cite{harris96} lists only three Galactic GCs that lie
anomalously above the bulk of the Galactic GCs in the plane $r_h$ vs
$M_V$, one of which is NGC~2419. The other two are M54 and $\omega$
Cen, the most luminous Galactic GC, and one with a large internal
spread in metallicity, also believed to be the stripped core of a
dwarf galaxy accreted by the Milky Way.  For these reasons,
\cite{vdb04} and \cite{mackey05}, among others, suggested that
NGC~2419 is also the remnant of an accreted dwarf galaxy.  In fact,
\citet{irwin99} and \citet{newberg03} suggested that NGC~2419 is one
of many GCs associated with the disrupted Sgr dSph, but \citet{law10}
firmly concluded that it is not.  Therefore, the cluster is potentially
a piece of an unidentified, tidally disrupted dSph.

To follow up on these earlier suggestions, \cite{deimos} obtained
moderate resolution spectra of a large sample of luminous giants in
NGC~2419 and analyzed line strengths in the region of the
near-infrared Ca triplet.  We found that there is a small but measurable
spread in Ca abundance of $\sim$0.3~dex within this massive metal-poor
globular cluster.  We present here an analysis of high dispersion
spectra of seven luminous giant members of NGC~2419, whose chemical
inventory we compare
to that inferred from a similar sized sample of luminous RGB members of the nearby
inner halo globular cluster M30 (NGC~7099), observed and analyzed
in the same way as the NGC~2419 stars. Note that
these two globular clusters
have the same metallicity in the 2003 version of the
on-line database of \cite{harris96}.
We explore two issues:

\begin{enumerate}

\item Is the chemical inventory in such a distant cluster identical to
  that of a globular cluster of similar [Fe/H] in the inner halo?

\item What star-to-star variations are revealed by our high-dispersion
spectra?

\end{enumerate}

\section{Observations and Analysis\label{section_obs} }

\begin{figure}
\epsscale{0.85}
\plotone{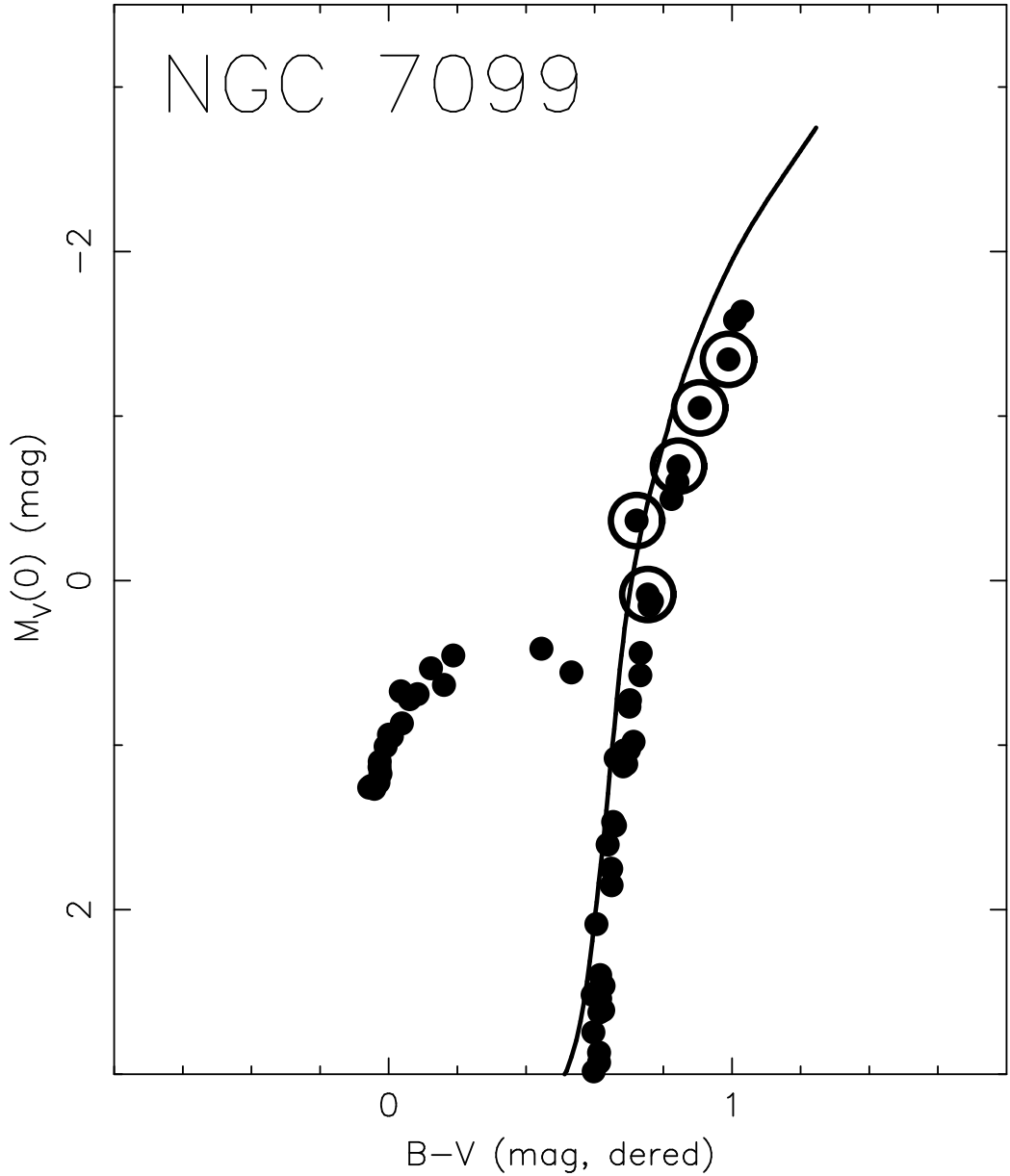}
\caption[]{A $M_V(0)~-~(B-V)_0$ CMD for NGC~7099 is shown from the
\cite{stetson05} on-line database.  The RGB stars in our HIRES
sample are circled.  This is a sparse cluster, and there are few
stars on the upper RGB. An $\alpha$-enhanced Y2 isochrone \citep{yale03}
with [Fe/H] $-2.2$ dex and age
of 12~Gyr is shown.  
\label{figure_7099_cmd}}
\end{figure}

\begin{figure}
\epsscale{0.85}
\plotone{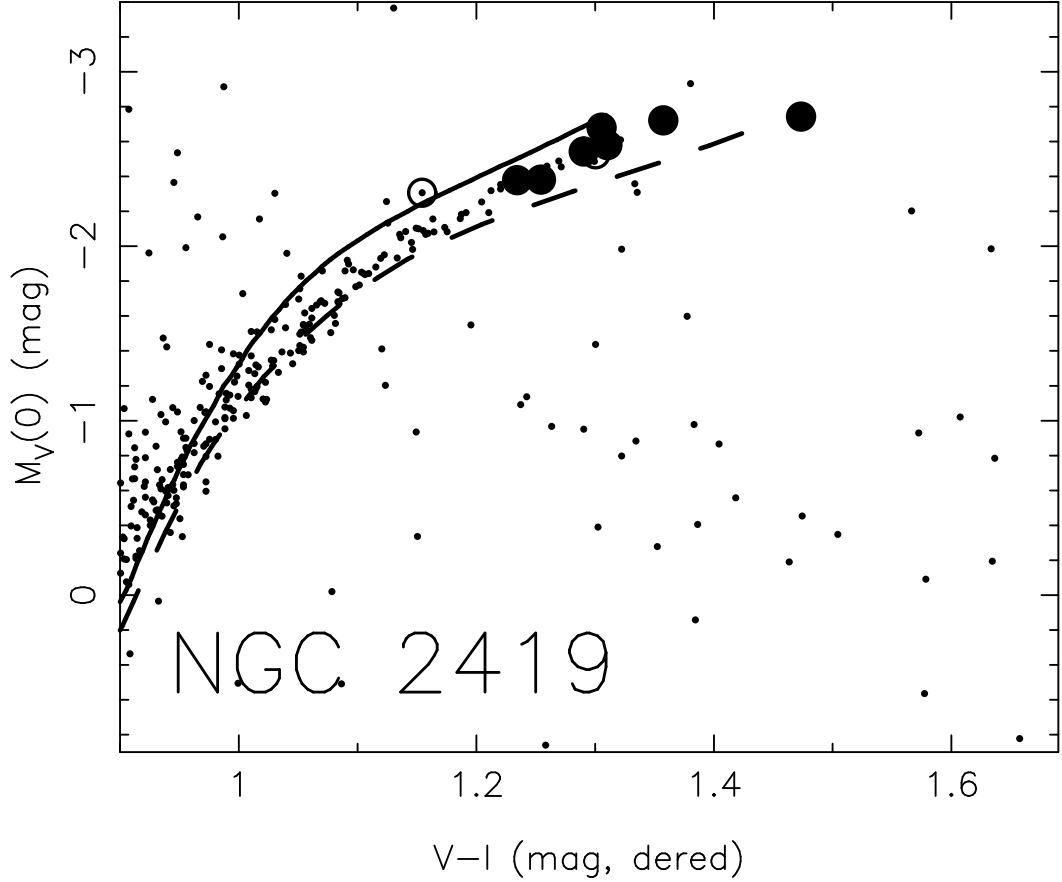}
\caption[]{A $M_V(0)~-~(V-I)_0$ CMD for the RGB in NGC~2419 is shown from the
\cite{stetson05} on-line database.  
The 7 giants in our HIRES sample are indicated by large filled circles.
 Two other stars with lower SNR HIRES
spectra, whose membership is questionable, are indicated by
open circles.  The two 12~Gyr Y2 isochrones \citep{yale03} 
shown have [Fe/H] $-2.2$ and $-1.9$~dex, with [$\alpha$/Fe] +0.3~dex. 
\label{figure_2419_cmd}}
\end{figure}

We obtained high-spectral-resolution and reasonably high-SNR spectra
of 7 luminous giants in NGC~2419 using HIRESr \citep{vogt94} on
the Keck~I 10~m Telescope.  
The stars were selected from both the early spectroscopic study of
\cite{suntzeff88} and the current version of  the
  on-line photometric database described by \cite{stetson05}.
The sample stars range in $V$ from 17.2 to
17.6~mag.  The spectral range is 3900 to 8350~\AA\ with  gaps
between the reddest echelle orders.  We used a 1.1~arcsec slit,
equivalent to 6.3 pixels at 15$\mu$/pixel, to achieve a spectral
resolution of 34,000.  The maximum total exposure time for one star
was 2.5 hours, split into shorter segments to improve cosmic ray
removal. The code MAKEE\footnote{MAKEE was developed
by T.A. Barlow specifically for reduction of Keck HIRES data.  It is
freely available on the world wide web at the
Keck Observatory home page, 
http:\\www2.keck.hawaii.edu/inst/hires/data\_reduction.html.}
was used to reduce the HIRES spectra, which
were obtained in various runs over the past 5
years.  The majority of the seven spectra have a SNR exceeding 90 per
spectral resolution element at 5500~\AA\ in the continuum.  The SNR
degrades towards the blue, and we often eliminated lines bluer than
5000~\AA\ for species with many detected features.  The radial
velocities for these seven stars have a mean of $-20.6$~km/sec with
$\sigma$ of 3.7~km/sec.  \cite{deimos}
discussed membership issues for NGC~2419 in detail.  
Our sample of 5 RGB
members in the sparse nearby GC M30 (NGC~7099) were observed
with the same HIRES configuration; all the resulting spectra
are of high SNR. Table~\ref{table_sample} gives details of the samples and
spectra.  Figs.~\ref{figure_7099_cmd} and \ref{figure_2419_cmd}
show CMDs for the full Stetson on-line database \citep{stetson05}
as well as our HIRES sample for each of these two GCs
with 12~Gyr isochrones from the Y2 grid \citep{yale03} superposed.


The determination of stellar parameters followed our earlier papers,
\cite[see, e.g.][~and references therein]{cohen05a}
and is based on $V-I$, $V-J$, and $V-K_s$ where the optical colors are from
\cite{stetson05} and the infrared colors are from 2MASS
\citep{2mass1,2mass2}. The uncertainties in 2MASS $K_s$ are
rather large for the faint NGC~2419 giants, and only two of the
NGC~7099 giants have an $I$~mag in the \cite{stetson05} database. 
We use the predicted color grid as a function of $T_{\rm eff}$,
log($g$), and [Fe/H]\footnote{The 
standard nomenclature is adopted; the abundance of
element $X$ is given by $\epsilon(X) = N(X)/N(H)$ on a scale where
$N(H) = 10^{12}$ H atoms.  Then
[X/H] = log[N(X)/N(H)] $-$ log[N(X)/N(H)]$_{\odot}$, and similarly
for [X/Fe].}  
from \cite{houdashelt00}.  If we use the recent $T_{\rm eff}$ --
color relatoins of \cite{irfm_teff}, which are not calibrated
for such luminous giants, we obtain $T_{\rm eff}$ $\sim$35~K
lower for the cooler luminosity stars in our sample for
NGC~2419, ranging up to 105~K lower for the most luminous
star in this GC.
The surface gravities
were calculated assuming a mass of 0.8~$M_{\odot}$, the known
distance, and the (low) interstellar extinction to each GC.  The isochrone of
the upper RGB for a very metal-poor old stellar system
is such that an error of 100~K in $T_{\rm eff}$ produces
an error of 0.2~dex in the inferred surface gravity using
this procedure.

The red giants in our sample in NGC~2419 span 0.36~mag
in $V$ and a range of less than 200~K in $T_{\rm eff}$.
The RGB stars in our sample for NGC~7099 are somewhat hotter
in the mean
due both to the somewhat lower metallicity of this GC 
and to the scarcity of upper RGB stars in this rather sparse
cluster.

The detailed abundance analysis for each of the cluster giants follows
J.~Cohen's previous work \citep[see, e.g.,][]{cohen05a} using
\cite{kurucz93} LTE plane-parallel 
model stellar atmospheres and the analysis code MOOG
\citep{moog}.  The adopted transition probabilities are largely
from NIST version 3. Note that the $gf$ values adopted here for
the observed Mg~I and Ca~I lines have
been updated from those we consistently used
up to the present \citep[see, e.g.][]{cohen05a}
to match the current on-line values in
the NIST version 4.0 \citep{nist} database. 
The  changes 
in abundance resulting from these updates are small, about
$-0.05$~dex for Mg for a typical set of observed lines,
  and  $+0.03$~dex for Ca~I
where more lines are usually observed,
only some of which have $gf$ values that have been updated.

Hyper-fine energy level splitting was used where
appropriate, with many such patterns taken from
\cite{prochaska}.  The adopted damping constants are those
described in \cite{cohen05a}.

Although it appears that small non-LTE corrections
should be applied for several species, for example Na~I \citep{takeda03}
and Ba~II \citep{nonlte_ba2}, for consistency with
our earlier work,
the only non-LTE correction we have
implemented is a fixed value of $+0.60$~dex for
the resonance Al~I doublet at 3944 and 3961~\AA\ following
\cite{al1_nonlte}, \citep[see also][]{andrievsky08}.
Non-LTE for K~I is discussed in \S\ref{section_s1131}.

Equivalent widths were measured in some cases by the script
described in {\S}3 of \cite{cohen05a},
which automatically
searches for absorption features, fits a Gaussian,
then matches those found
to a master line list given the radial velocity of the star.
For other sample stars,
W.~Huang used an IDL script to determine $W_{\lambda}$
for the set of lines in the master list.
There was extensive hand checking
of weak features, of most rare earth lines, and of many
of the strongest accepted lines. Lines with $W_{\lambda} > 175$~m\AA\
were rejected unless the species  had only a few detected features.
The resulting values, together with the
atomic parameters for each line used, are given
in Tables~\ref{table_2419_eqw} and \ref{table_7099_eqw}.

The  $W_{\lambda}$ for the  3961~\AA\ Al~I line, when given, 
are particularly uncertain for the NGC~2419 giants; the
3944~\AA\ line was never used to due contamination by CH.
Both lines are uncomfortably far in the blue,
where the SNR of the HIRES spectra for such faint stars
is degraded, but the 3961~\AA\ line is a key feature for this element as 
the 6690~\AA\ Al~I doublet is very weak and difficult to
detect at the low metallicity of this GC.

Because of the uncertainty of the photometry for the NGC~2419 giants, 
we felt free to make
small adjustments to the photometrically derived stellar parameters,
primarily changing $T_{\rm eff}$.  These adjustments,
detailed in Table~\ref{table_temp}, improved the
excitation equilibrium for \ion{Fe}{1} lines and the ionization balance
between neutral and singly ionized lines of Fe and Ti.  No such
adjustment exceeded 100~K, and was less than
50~K for four of the NGC~2419 giants.  In the case of M30,
only one star had $T_{\rm eff}$ modified by more than 20~K.
For each star in NGC~2419, we measured 67 to 108 \ion{Fe}{1}
lines, which allowed us to determine the microturbulent velocity
directly from the spectra; every star in our sample in NGC~7099
had 99 or more detected Fe~I lines.

Table~\ref{table_fe_slopes} gives the slope of a linear
fit to the Fe abundance deduced from each of the Fe~I lines
detected in a specific sample star
as a function of excitation potential, reduced equivalent
width ($W_{\lambda}/{\lambda}$), and $\lambda$, where the
first is most sensitive to the adopted $T_{\rm eff}$, the
second to $v_t$, and the third indicates any systematic
problems in continuum opacity as a function of wavelength.
Values are given for both GCs.  In the ideal case, these slopes
are zero.  For NGC~2419 the typical
range in $\chi$ for the Fe~I lines is 0.9 to 4.2~eV;
the stars with the best spectra span a larger
range from 0.0 to 4.8~eV.  The wavelength range for Fe~I lines
in most of the NGC~2419 (NGC~7099) spectra is $\sim$2400~\AA\
($\sim$3000~\AA),
and a typical range in reduced equivalent width for
such spectra is $-5.4$ to $-4.5$~dex for NGC~2419
and $-6.3$ to $-4.4$~dex for NGC~7099. 
The crucial slope with $\chi$ ranged from $-0.083$
to $-0.038$~dex/eV for the NGC~2419 sample, and is
$-0.048\pm{0.007}$~dex/eV for the M30 RGB sample.

The resulting abundances for the 7 RGB stars in NGC~2419
and the 5 in M30 are given in
Tables~\ref{table_ngc2419_abund} and  \ref{table_7099_abund},
where all abundances are relative to Fe~I.
The adopted Solar absolute abundances for each element can be
inferred from the entries in Table~\ref{table_7099_abund}.
With regard to the ionization equilibrium, that for Fe~I
vs Fe~II is generally quite good, within 0.1~dex, for all
stars in both NGC~2419 and in M30. 
The ionization equilibrium of Ti~I vs Ti~II is somewhat worse, typically
within $\sim$0.2~dex, with the Ti~II lines giving a somewhat
higher Ti abundance.

Tables~\ref{table_abund_sens_abs} and \ref{table_abund_sens_rel}
give the sensitivity of the derived
absolute abundances [X/H] and those relative to Fe [X/Fe]
for small variations in the various relevant stellar and
analysis parameters.

\section{Alternative Choices of Stellar 
    Parameters \label{section_caspread} }

The choice of stellar parameters is crucial for determining
absolute stellar abundances, although less critical for abundance
ratios [X/Fe] as many of the systematic effects cancel,
as is shown by comparing the entries for each species in
Table~\ref{table_abund_sens_abs} with those in 
Table~\ref{table_abund_sens_rel}.  We have
therefore explored several methods of determining $T_{\rm eff}$.
We call the $T_{\rm eff}$ determined by the mean of that
inferred from each available
color ($V-I,~V-J,$ and $V-K_s$), the photometric $T_{\rm eff}$.  
The value of $T_{\rm eff}$ adopted for the abundance analysis,
 $T_{\rm eff}$(adopt),  is initially
set to $T_{\rm eff}$(phot), with small adjustments permitted to improve
the Fe~I slopes and the ionization equilibrium as described above.

Another
$T_{\rm eff}$ can be determined by assuming that all the stars are
RGB members of their cluster, and that all the cluster stars
have the same initial chemical inventory and age.  Hence they must 
lie along a single old very metal-poor isochrone.
Here we are using a single magnitude,
chosen on the basis of its high measurement accuracy
and discrimination along the isochrone,
to determine $T_{\rm eff}$(iso).
This removes the issue of random errors in colors,
which is of particular concern  for NGC~2419, with its large distance and
hence faint RGB stars near the limit for 2MASS photometry.
With a large enough sample, the ``isochrone''
itself can be defined from the mean locus of the stars in the cluster CMD,
as was done in \cite{cohen05a}, see also the large VLT
study of \cite{carretta10} and references therein.
In the present case, even if the HIRES sample in NGC~2419
were much larger, 
the concern
that this particular GC may not be chemically uniform would 
limit the applicability of such an approach.
If the isochrone chosen is appropriate
for the cluster,
then the mean ${\Delta}[T_{\rm eff}(phot) - T_{\rm eff}(iso)]$ will be zero.
If the isochrone is close, but not exactly that of the cluster, then
the mean difference will be a constant.

Table~\ref{table_temp} gives the values of the adopted, photometric,
and isochrone $T_{\rm eff}$ for the sample of RGB stars in NGC~2419.
The rms $\sigma$ about the mean for the set of available colors
from the three we consider is
given for the $T_{\rm eff}$(phot) in parentheses. 
The photometric $T_{\rm eff}$ was  adopted for the detailed
abundance analysis  for 3 of the 5
stars in NGC~7099.  For the other two, the differences are 18~K (0.8 $\sigma$)
for S31 and 79~K (1.2 $\sigma$) for S34.  The isochrone and photometric
values of $T_{\rm eff}$ for this nearby GC are essentially identical.
For NGC~2419 that is not the case. The brightest star (S223)
at the RGB tip is somewhat redder than would be expected from the isochrone.

\section{Inner vs Outer Halo }

\begin{figure}
\epsscale{0.9}
\plotone{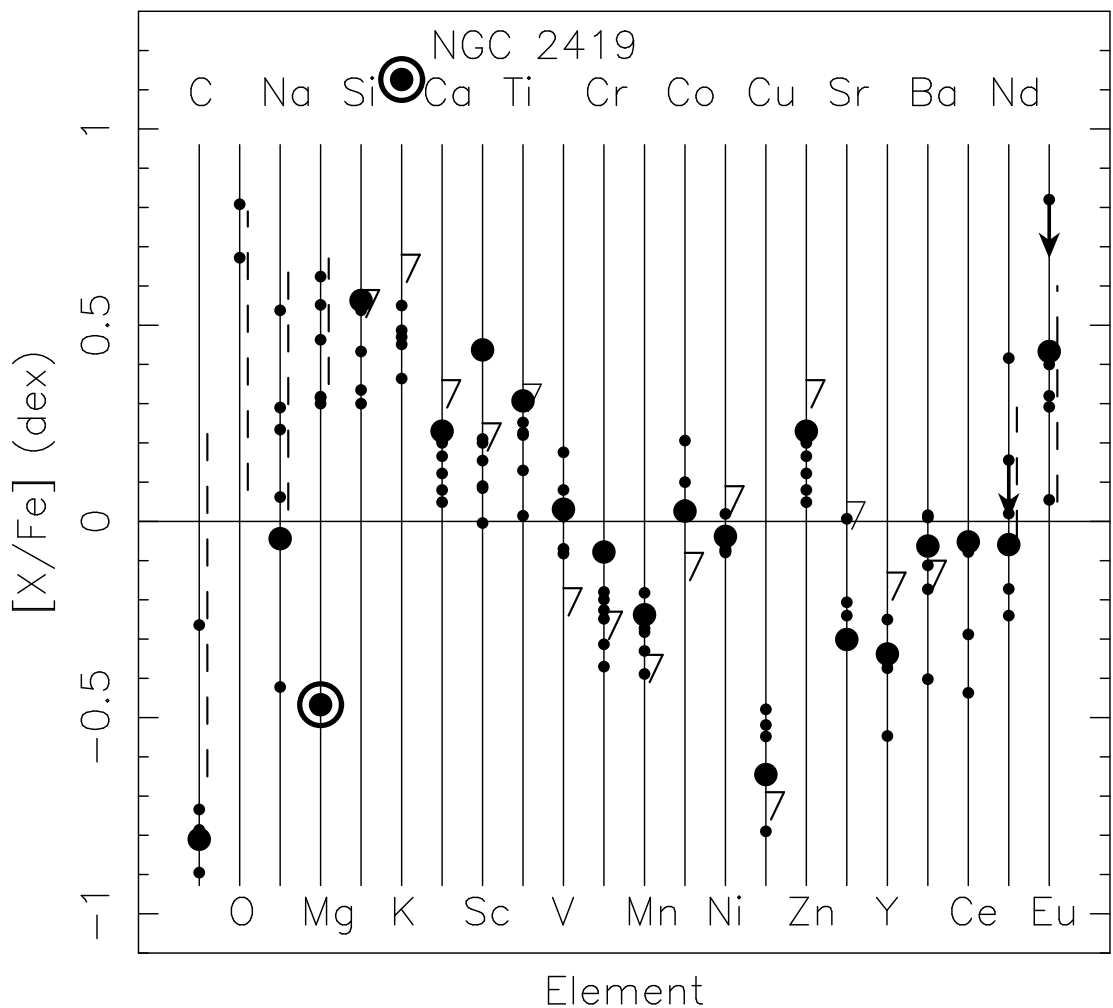}
\caption[]{Abundance ratios for the detected elements in the NGC~2419
sample of 7 red giants, with the large filled circles indicating
S1131. The highly unusual abundance ratios
of [Mg/Fe] and [K/Fe] for NGC~2419 S1131 are
circled.  The reference species is FeI for neutral
species and FeII for ionized ones.  
The averages of TiI and TiII and of SrI and SrII 
are plotted  when both are available. 
Average abundance ratios from  our analyses of HIRESr spectra
of red giants in the inner halo cluster M30 (NGC~7099) are indicated
by ``7''.  For elements with a large range in M30, the range is
indicated by a dashed line.
\label{figure_abund}}
\end{figure}

A comparison of the mean abundance ratios in the two globular
clusters discussed here
M30  (NGC~7099), at a distance of only 8~kpc, with the very distant
outer halo
globular cluster NGC~2419 is given in Table~\ref{table_abund_mean}
and illustrated in detail in Fig.~\ref{figure_abund}.
There is no substantive difference between the elemental abundance
ratios for luminous red giants in the outer halo cluster NGC~2419 and
the ratios of inner halo clusters.   The inner halo abundance pattern 
is also
observed in other distant clusters: NGC~7492
\citep[$R_G \sim 25$~kpc,][]{cohen05b}, the low luminosity GC Pal~3
\citep[$R_G \sim 90$~kpc,][]{pal3}, and in coadditions of
high-resolution, low-SNR spectra of 19 luminous giants in the very
distant, low-luminosity GC Pal~4 \citep[$R_G \sim 109$~kpc,][]{pal4}.
The low [C/Fe] ratios seen among the NGC~2419 and M30 
luminous giants are common
among such stars in inner halo GCs \citep[see, e.g.][]{m13c}, 
presumably due to depletion via  deep mixing.

When comparing Galactic GC abundances to those of dSph
satellites of the Galaxy, it has become apparent
via extensive surveys in recent years 
\citep{cohen_draco, cohen_umi} and at moderate resolution
the extensive work of \cite{kirby4} that
the metal-poor end of all known systems seems to
have, at least to first order, an identical chemical 
inventory.  That should not be a surprise, since differences
produced through different star formation rates
or gas flow history (accretion and/or outflows)  between the dSph satellites and
the Galactic GCs will only show up at late times.  Initially
only the metals produced in and ejected by SNII
enrich the system's gas, and hence,
except for small metallicity dependent effects
on nucleosynthetic yields, all
very metal-poor stellar systems should have identical abundance
ratios.  Those GCs with peculiar abundance ratios,
including no enhancement of the $\alpha$-elements, have
intermediate [Fe/H] and most are
known to be associated with the ongoing accretion of the Sgr dSph;
examples include
Pal~12 \citep{cohen_pal12} with [Fe/H] $\sim -0.7$~dex, 
Terzan~7 \citep{terzan7} with
[Fe/H] $-0.6$~dex, as well as
Rup~106 \citep{rup106} with [Fe/H] $-1.45$~dex, whose age is several
Gyr less than the bulk of the halo, suggesting an accretion origin.

\section{Comparison with Previously Published Abundance Analyses}

The only previously published high-dispersion analysis of any star in
NGC~2419 is for the star S1305, which was observed by
\cite{shetrone01}\footnote{\cite{shetrone01} call this star RH~10.
The coordinates of this star, privately communicated from
  M.~Shetrone, match those of S1305.}.  Our measurement of [Fe/H] is
0.13~dex higher than the value they obtained, which we ascribe
largely to our adopted $T_{\rm eff}$ being 125~K
higher than theirs.  The two sets of [X/Fe] for all elements in 
common through atomic number 30 (Zn) are in good agreement
except for Na.  Na is a difficult case as the NaD lines are somewhat
corrupted by interstellar features at the  cluster's $v_r$,
while the 5682, 5688~\AA\ doublet used by \cite{shetrone01}
is very weak.  Among the heavier elements detected in
both studies, only Y (in the form of Y~II)
has a large discrepancy.  There are 3 detected lines from
our spectra, and a claim of 4 from theirs, but the line 
at 4900.1~\AA\ is quite crowded, and perhaps should be
ignored.  For the two Y~II lines in common, our $W_{\lambda}$
are $\sim$20\% smaller.  Considering how faint this star is 
($V = 17.61$~mag) we regard the overall agreement
in abundances between the two analyses to be very good.

\cite{shetrone03} observed one star in NGC~7099.  Once
the difference in adopted solar Fe abundance is taken into
account, their derived [Fe/H] for this star differs from
the mean for our sample of five stars by only 0.05~dex.
The abundance ratios agree satisfactorily in most cases,
with 7 elements differing in [X/Fe] by less than 0.10~dex.
Their UVES observations  cover only 4800 to 6800~\AA,
and so are missing many key blue lines for the rare earths,
where the differences in relative abundances between
the two studies tend to be the largest. The choice of
adopted transition probabilities, particularly for the rare earths, also contributes;
we adopted them from the recent studies of Lawler
and collaborators, 
\citep[see, e.g.][for Eu as an example]{lawler01} 
when available, while \cite{shetrone03} tended to use older values.

The Padua group (see, e.g. Carretta et al 2009abc and
Carretta et al 2010) has completed a large VLT study
of the Na/O anti-correlation in GCs.  Although their survey does not
include NGC~2419, they   have a observed a
sample of 10 RGB stars in M30 with UVES.  Thus far they have  published
abundances for only the light elements.  
Table~\ref{table_7099_comp} presents a comparison of our
results from HIRES at Keck with their UVES/VLT study
and highlights  the good agreement between
these two completely independent detailed abundance
analyses for RGB stars in M30.


\section{Consistency with Our Previous Deimos Range in Ca Abundance 
\label{section_ca_2419} }

\begin{figure}
\epsscale{0.9}
\plotone{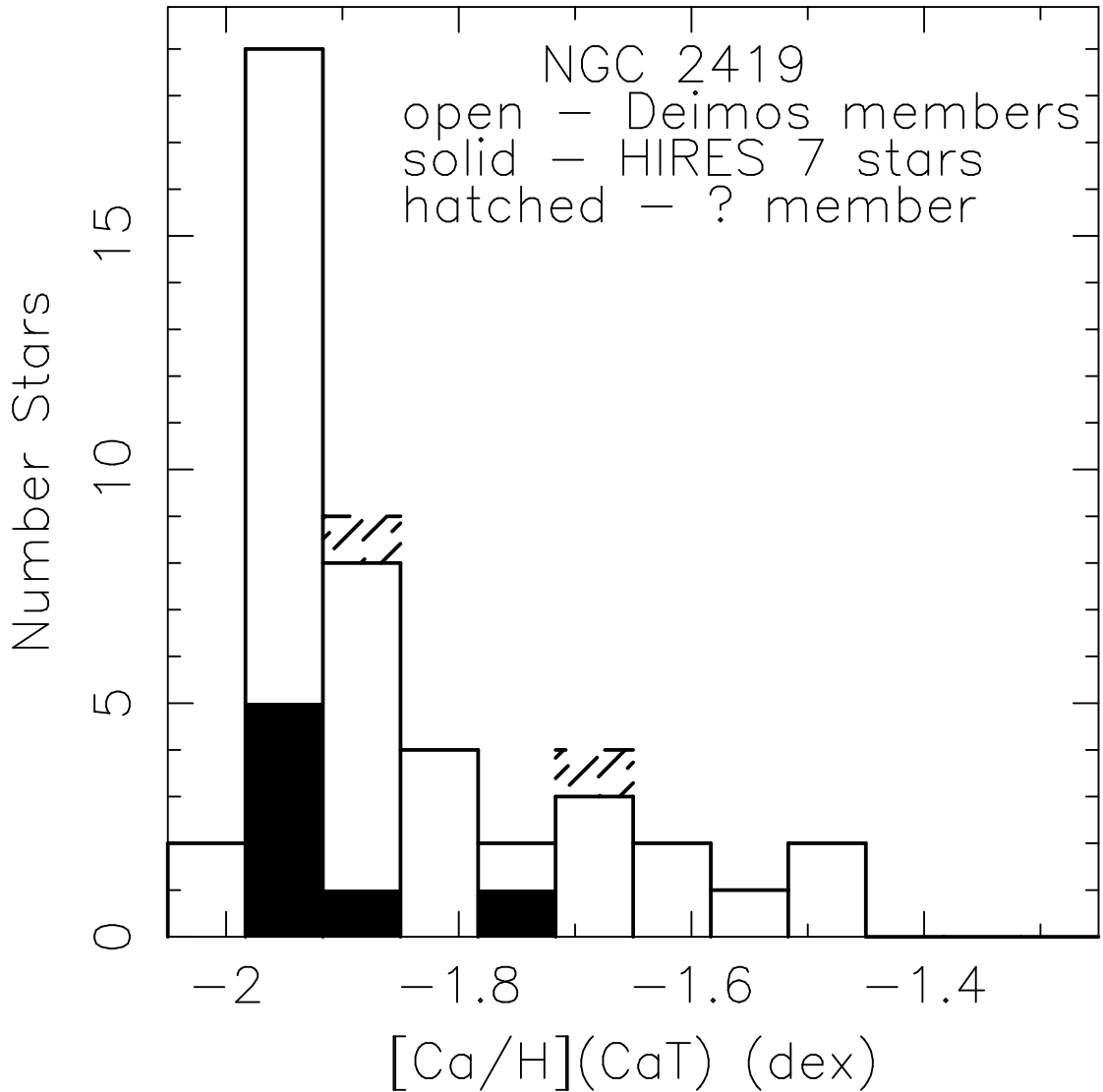}
\caption[]{A histogram of [Ca/H]  
as inferred from the Deimos moderate
resolution spectra of \cite{deimos} is shown for the sample of 43 definite members of NGC~2419
isolated in that paper.  The present sample of 7 RGB stars in this GC with HIRES
spectra is shown by the solid fill.  The two 
additional stars in the Deimos survey
which are probable members, both of which have low SNR HIRES spectra, are indicated
by the hatched areas, which are placed above the histogram defined by
the definite members.
\label{figure_cah_hist}}
\end{figure}

Our previous medium-resolution study of RGB stars in NGC~2419
\citep{deimos} uncovered a range in [Ca/H] of a factor of 
two or three among the sample of 43 stars found to be definite
members of this GC.  The range in $V$ of these stars is from 17.4 to 19.3~mag.
Fig.~\ref{figure_cah_hist} shows a histogram of [Ca/H] for these
stars, on which is superposed  the distribution of the
giants in NGC~2419 with  high quality HIRES spectra analyzed 
here.
The two additional probable members discussed in 
\cite{deimos}, each of which has
some remaining concern about membership, are indicated in this figure
with cross hatching.

The first point to note is that the histogram of [Ca/H] for an unbiased
sample of members of NGC~2419, the moderate resolution Deimos survey
of  \cite{deimos}, is sharply peaked at  [Ca/H] $-1.95$~dex.
While there is a rapidly declining tail extending towards higher metallicities,
the low metallicity peak dominates, and the fraction of stars
in the extended tail is small.  The minimum  [Ca/H] is $-2.0$~dex.
Only 13 (30\%) of the 43 definite members lie outside the range
$-2.0$ to $-1.85$~dex.

Our sample of RGB stars in NGC~2419 with HIRES
spectra good enough for a detailed abundance analysis 
contains only 7 stars.
This sample was, to a large extent, selected by brightness before
the full analysis of the moderate resolution spectra was completed.
One of these 7 stars (S1131) has
[Ca/H] $-1.72$~dex from its HIRES spectrum ($-1.76$~dex from
its Deimos Ca triplet region analysis) and a second star (S973) is
marginally above the $-1.85$~dex [Ca/H] cutoff at
$-1.81$~dex from our detailed abundance analysis
($-1.93$~dex from its moderate resolution Deimos spectrum).
The others all are more metal-poor than [Ca/H] $= -1.85$~dex as deduced
from both their HIRES and Deimos spectra.

The total range in Ca and Fe abundance  within the NGC~2419 sample
is small (see Table~\ref{table_ngc2419_abund}.
We used the three different choices of $T_{\rm eff}$
discussed in \S\ref{section_obs}
and given in Table~\ref{table_temp} as an indication of the maximum
range in this parameter that might be appropriate for each of the NGC~2419 
stars\footnote{We adjusted log($g$) appropriately in each case to maintain the star
on the RGB.}.
In effect, we carried out the abundance analysis three times for each star
using different sets of adopted stellar parameters.
We find that for each of these three choices, S1131 {\it{always}} has
the highest value of [Ca/H] of the 7 stars in the HIRES sample for NGC~2419,
 and it is ${\sim}0.2$~dex (2 to 3 $\sigma$) higher than
the mean value for the other 6 stars in this cluster.
The Fe abundance for S1131, as derived from either Fe~I or Fe~II,
is always the highest or second highest as well, but is
within 1~$\sigma$ of the mean of the other cluster stars. 

The uncertainties
in  absolute abundances (see Table~\ref{table_abund_sens_abs})
are daunting for such a small range in Ca abundance in such
a distant cluster with spectra and near-IR photometry
that are less than ideal.  Lowering 
the $T_{\rm eff}$ of S1131 in NGC~2419 by $\sim$100~K
relative to the other RGB stars in this cluster in our HIRES
sample would eliminate its high Ca and Fe abundance.
But this  requires a change 
of this one star relative to all the others which
is about half of the
entire range in $T_{\rm eff}$ spanned by the 7 cluster giants,
for which there is no justification.
Given the small size of our HIRES sample,
all we can say is that the Ca/H distribution of the HIRES
sample in NGC~2419 is consistent with that of the Deimos sample, which
does appear to show a small but measurable range in Ca/H among
the cluster members \citep{deimos}.

Both of two probable additional members of NGC~2419 discussed
in \cite{deimos} from the Deimos sample, shown by cross hatching in
Fig.~\ref{figure_cah_hist},
have been observed with HIRES.  However observations were
terminated after the first 30~min exposure revealed
concerns about cluster membership.  Hence
the resulting HIRES spectra for these two stars are not good enough for a detailed
abundance analysis.
One of these (S951) has a spectrum which appears to be that of a RGB
star in this very metal-poor GC, but with  $v_r$ off from the cluster mean
by $\sim$15~km s$^{-1}$.  This star has a value from an analysis of
the Ca triplet lines in its Deimos spectrum of
[Ca/H] of $-1.91$~dex, consistent with the bulk of the cluster
population.  The concern with the second star, S1673,  is twofold; it lies
slightly bluer than the main RGB locus of NGC~2419 in $V-I$
(see Fig.~\ref{figure_2419_cmd}) and its  HIRES spectrum
appears metal-rich.  The [Ca/H] inferred from its Deimos 
spectrum ( $-1.68$~dex)
puts this star, if it is a member, firmly in the metal-rich
tail of the histogram shown in Fig.~\ref{figure_cah_hist}.  
See \cite{deimos} for further discussion
regarding membership in NGC~2419 of these two stars.

There are 13 stars in the Deimos sample that may
be members of NGC~2419 with $V < 17.7$, which is slightly fainter
than the faintest star in the present HIRES sample analyzed here.
In addition to all 7 RGB stars in our present HIRES sample,
both
of the two probable cluster members discussed above
are brighter than that cutoff.
This leaves 4 stars brighter than $V = 17.7$~mag without
HIRES observations at present,
3 of which have [Ca/H] $\sim -1.9$~dex. Only one is in the high
metallicity tail for NGC~2419.  

Our present
HIRES sample, which was selected primarily on the
basis of brightness given the 84~kpc distance to NGC~2419, 
is not ideal for exploring the star-to-star variation.
The small variations of [Ca/H] or [Fe/H] from the 7 stars with
high dispersion spectra are only marginally larger than the
uncertainties, but are broadly consistent with our
Deimos results \citep{deimos} which found a range
in [Ca/H] within NGC~2419.  In the future we hope to provide a sample that better
probes the metal-rich tail of  NGC~2419  as
defined by the larger Deimos sample of \cite{deimos}.


\section{One Star with a Peculiar Abundance Pattern \label{section_s1131} }

One star, NGC~2419 S1131,deviates from the
archetypal GC abundance pattern. 
 S1131 is
a definite member of NGC~2419 based on its 
HIRES radial velocity (its heliocentric $v_r$ is $-17.2$~km s$^{-1}$)
and spectrum; see also \cite{deimos}.  As discussed above
in \S\ref{section_ca_2419}
our detailed abundance analysis
suggests that this star is slightly more metal-rich 
than the other NGC~2419 giants
in Fe and Ca as well as 
several other elements with many detected lines.
It has  an
enhancement of $\sim 0.2$~dex in [Ca/H] 
(with slight variation depending on the choice of
$T_{\rm eff}$ discussed in \S\ref{section_caspread}), corresponding
to Ca enhanced by  a factor of 1.6, compared to the mean 
of the other 6 NGC~2419 giants.

It is interesting to note that S1131 is the only one of the six
NGC~2419 sample giants included in the moderate resolution study of
\cite{deimos} found to be Ca-rich.  The remaining 6 NGC~2419 giants in
the HIRES sample  all lie at the low Ca
abundance end of the distribution inferred by \cite{deimos}. 
However, given the small abundance
range involved, observations of additional giants believed
to be more metal-rich than the bulk of the cluster stars
are required for a definitive confirmation for star-to-star
variation of Ca, Fe, and other heavy elements within
NGC~2419.

\begin{figure}
\epsscale{0.83}
\plotone{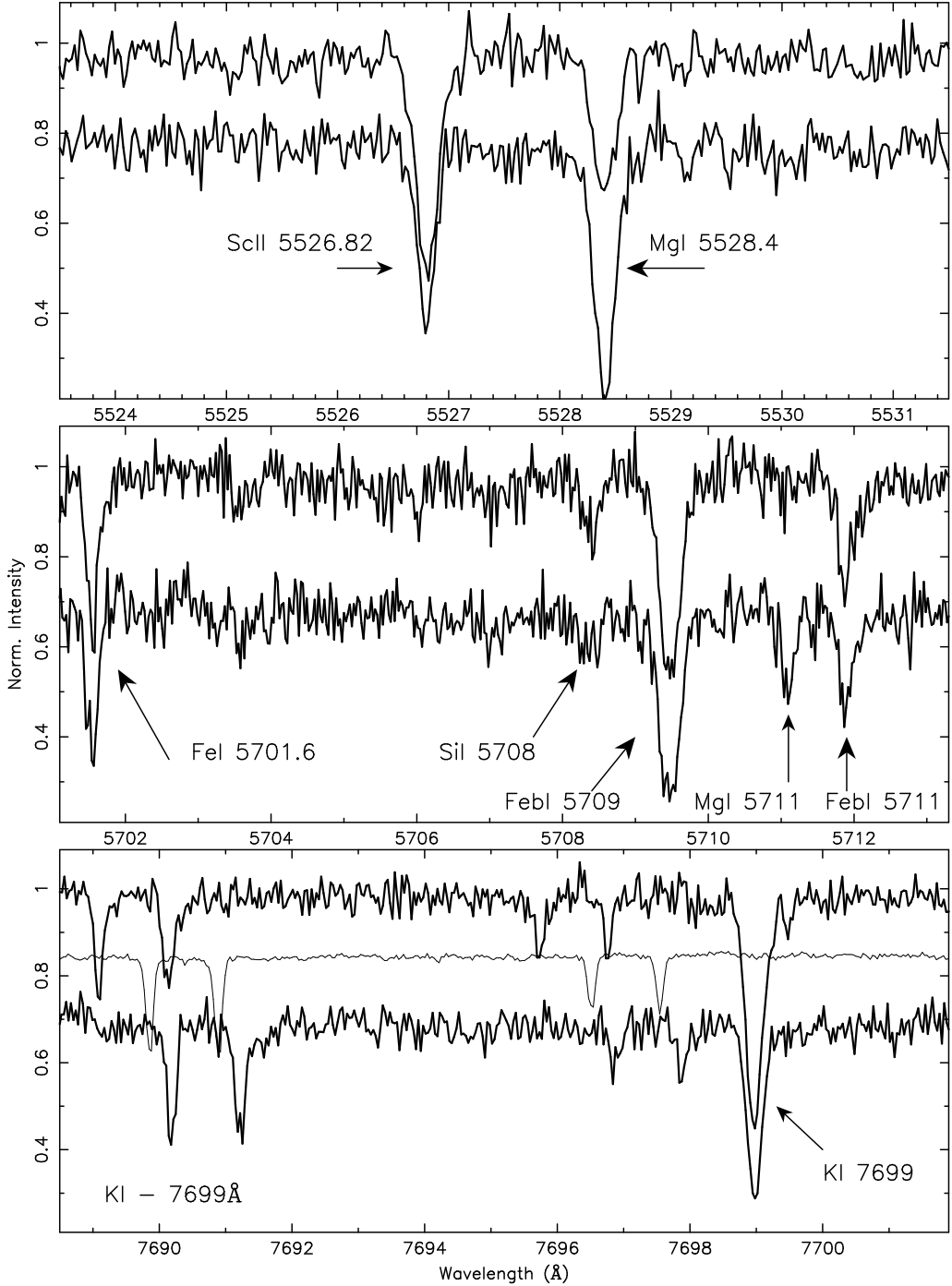}
\caption[]{The regions of the spectrum near two Mg~I lines
and a line of K~I, shifted in wavelength into
the rest frame, are shown for NGC~2419 S1131 as compared
to S1209, whose spectrum is  shifted slightly lower for clarity.
These two giants differ in $V$ by 0.2~mag and in deduced
$T_{\rm eff}$ by less than 100~K. 
The lines identified include a number of blends of FeI lines
or FeI and NiI lines labeled as ``Febl''.  S1131 
is more metal-rich and Mg-poor than S1209 and the other
NGC~2419 giants.  The thin line in the bottom panel is that
of a rapidly rotating B star.  The slight shifts in wavelength
of the telluric lines between the spectra is a reflection
of the differing heliocentric corrections for each spectrum.
\label{figure_spectra}}
\end{figure}

Furthermore, NGC~2419 S1131 has some anomalous abundance ratios,
irrespective of whether its $T_{\rm eff}$ has been overestimated.  It
has [Mg/Fe] = $-0.47$~dex from five \ion{Mg}{1} lines, in contrast
to the mean of the other six stars,
$+0.44\pm0.06$~dex.  In deriving the Mg abundance for S1131, the upper limit
to the strength of the 5711~\AA\ line was considered as a detection,
and the two unblended Mg triplet lines were retained.  These two lines
give Mg abundances slightly higher than the mean, but the highest
Mg abundance is from the 5711~\AA\ line, 0.18~dex above the mean
of the 5 lines used.  Note that the first ionization potential of
Mg is only 0.25~eV lower than that of Fe.  The [Mg/Fe]
measurement would increase by only 0.08~dex if $T_{\rm eff}$ were
reduced by 100~K (see Table~\ref{table_abund_sens_rel}).  
The regions of the spectra of two \ion{Mg}{1} lines
in this star and in S1209, also known as Suntzeff~16
\citep{suntzeff88}, with $T_{\rm eff}$ lower than that of S1131 by
only 85~K, are shown in Fig.~\ref{figure_spectra}.  Although most GCs
show a small spread in [Mg/Fe] \citep{gratton_araa}, 
generally interpreted as the result of proton-burning
of Mg into Al in
intermediate mass AGB stars,the expected enhancement of Al does not 
appear to be present in S1131.  The abundance pattern of this
specific star cannot be reproduced by proton-burning.
Furthermore  $-0.47$~dex is an
unprecedentedly low [Mg/Fe] abundance for a globular cluster star; the
only stars that have such low [Mg/Fe] ratios are the halo star
CS22966-043 \citep{ivans03}, a hot SX~Phe variable, 
and stars in dSph satellite galaxies, such
as COS~171 in Ursa Minor \citep{cohen_umi} and S58 in Sextans
\citep{shetrone01}.   NGC~2419
S1131 is unique in that no other elements appear depleted, including
Si.  A very low value of Mg and super-solar values of other alpha
elements are inconsistent with any Type~II SN yields
\citep[e.g.,][]{woosleyweaver,nomoto06,heger10}.

In addition, NGC~2419 S1131 also has an unusually high [K/Fe] of
$+1.13$~dex, also
illustrated in Fig.~\ref{figure_spectra}.   This figure includes
the spectrum of a rapidly rotating B star to demonstrate that
any telluric absorption at the wavelength of the K~I line,
taking into account the $v_r$ of this GC, is negligible.
Furthermore, two other NGC~2419 sample stars were observed
on the same run (see Table~\ref{table_sample}) and do not
show this abnormality.
This abundance ratio would
decrease by less than  0.10~dex if $T_{\rm eff}$ were reduced by 200~K. 
This is an internal comparison within our NGC~2419
  sample, which has a total range in $T_{\rm eff}$ of less than 200~K.
Studies of non-LTE corrections for the 7699~\AA\ 
resonance line of K~I
(the only line used here) by \cite{andrievsky10} and by \cite{takeda_k_nonlte}
 suggest that the appropriate non-LTE corrections
change by only 0.15~dex over such a small range in $T_{\rm eff}$,
ranging from about $-0.35$ to $-0.2$~dex.
Thus differential non-LTE corrections for K~I within our HIRES 
sample of luminous RGB stars  are much too small to explain the
deviant potassium abundance of S1131.

\cite{takeda_k_nonlte} very recently found a red giant in M13 and another
in M4 with similarly high [K/Fe].  They ascribe the unexpectedly
strong K~I resonance lines to an increase in activity or turbulent
velocity fields high in the stellar atmosphere where 
the core of such a strong line is formed.
However, while S1131 does show emission on both the red and blue wings
of H$\alpha$, the emission wings of H$\alpha$ 
in the spectrum of NGC~2419 S223 are considerably stronger,
yet this star does not show an anomalously strong K~I line.
Furthermore if this explanation is correct, 
other strong resonance lines arising from neutral species
of elements with (low) first ionization potentials comparable 
to that of K should be similarly affected.  Unfortunately
the NaD lines, the most likely candidates to check, are
badly corrupted
by interstellar absorption due to the radial velocity of NGC~2419.

A careful inspection of Fig.~\ref{figure_spectra} 
strongly suggests, by comparison with the spectrum of NGC~2419
S1209, which has a $V$ magnitude only 0.2~mag brighter than S1131, 
that S1131 is indeed
slightly more metal-rich in other species as well as K. 

The Li~I blend at 6707~\AA\ was too weak to be detected
in any of the NGC~2419 stars.  The upper limit for the equivalent
width of this feature of 7~m\AA\ in
S1131 corresponds to log[$\epsilon$(Li)]
of $-0.01$~dex, indicates extensive depletion from the Li production 
in the Big Bang, typical of red giants, which have extensive
surface convection zones.

The stellar parameters, spectra, and
analyses have been carefully checked and these specific differences in
abundance ratios discussed here (i.e. for Mg and for K)  
within the NGC~2419 sample are real. 


\section{Summary}

Our key result from the analysis of 7 luminous giant members of the
distant outer halo globular cluster NGC~2419 is that one of these
stars is extremely peculiar, having a very low [Mg/Fe] ratio, but
being normal in all other element ratios except for a high [K/Fe].
Similarly low [Mg/Fe] ratios are seen in stars in dSph satellites of the
Galaxy, specifically in Ursa Minor \citep{cohen_umi} and in Sextans
\citep{shetrone01}.  But in both cases these stars have low values of
other $\alpha$ elements, which is not the case for NGC~2419 S1131.
J.~Cohen has examined and analyzed spectra of $\gtrsim$100 stars in
Galactic GCs, and has never seen one similar to S1131 in NGC~2419.

Furthermore, the HIRES spectra support the small range in
metallicity within the giants in NGC~2419 found by \cite{deimos},
which might suggest that this distant outer halo GC is
the remnant of an accreted dwarf galaxy. A definitive result
will require further observations of the faint upper RGB
stars in this GC, concentrating on those whose moderate
resolution Deimos spectra from \cite{deimos}
suggest they are more metal-rich than
the bulk of the stars in this cluster, which will be carried out
in  a future campaign.

Ignoring the limited peculiarities of S1131, we find that there is
no substantive difference between the mean behavior of the abundance
ratios for six other red giants in NGC~2419 and those ratios characteristic of
the inner halo despite the structural peculiarities of NGC~2419. 
The influence of star formation rates
and other aspects of the detailed chemical evolution of a stellar
system only affect such ratios once  star formation
has been underway long enough for sources other than SNII to
become important contributors, which does not seem to be the case
in very metal-poor systems such as NGC~2419, nor in the
most metal-poor stars in the dSph satellites of the Milky Way
(Cohen \& Huang 2009, 2010; Kirby et al 2011).

Thus, like the previous studies based on detailed
abundance analyses of individual or co-added stellar spectra of stars
in distant outer halo clusters, the chemical inventory of the globular
cluster system appears to be independent of location.  With the
exception of those clusters known to be associated with the
currently disrupting Sgr dSph galaxy, every globular cluster
exhibits the same trends as a function of overall cluster
metallicity---parameterized by [Fe/H]---irrespective of kinematics or
location within the halo.  

The existence of a metallicity gradient in the Galactic halo is
controversial.  \citet{carollo07} found evidence for a break in the
metallicity distribution of halo field stars in the sense that the
more distant stars are more metal-poor;  this conclusion is
supported by \cite{rix10}.  Furthermore, \cite{carollo07} showed that
the metallicity within the outer halo population displays a gradient
toward even lower metallicities for the most distant stars or stars
showing the highest magnitude of rotation retrograde to the Milky Way
disk.  However, both \citet{sesar11} and \citet{schoenrich10} improved
upon \citeauthor{carollo07}'s study.  These new works,
which are disputed by \cite{beers11}, found no
evidence for a dichotomous halo, particularly in regard to the
kinematics and density profile.  Our measurements for NGC~2419 
additionally provide
evidence against a dichotomy in detailed abundances.  The outer halo
globular clusters show the same bulk abundance
properties---and, by extension, (short) star formation timescales---as the
inner halo globular clusters.

\acknowledgements

We are grateful to the many people who have worked to make the Keck
Telescope and its instruments a reality and to operate and maintain
the Keck Observatory.  The authors wish to extend special thanks to
those of Hawaiian ancestry on whose sacred mountain we are privileged
to be guests.  Without their generous hospitality, none of the
observations presented herein would have been possible.  We thank 
Alex Heger and Ken Nomoto for helpful conversations.  J.G.C.
and W.H. thank NSF grants AST-0507219 and
AST-0908139 for partial support. Work by E.N.K. was supported by NASA
through Hubble Fellowship grand HST-HF-01233.01 awarded to ENK by the
Space Telescope Science institute, which is operated by the
Association of Universities for Research in Astronomy, Inc., for NASA,
under contract NAS 5-26555.

{}

\begin{deluxetable}{l l rrrrr rr}
\tablenum{1}
\tablewidth{0pt}
\tablecaption{Data for Red Giant Members of NGC~2419 and of NGC~7099 With 
HIRES/Keck Spectra
\label{table_sample} }
\tablehead{
\colhead{Name\tablenotemark{a}} & \colhead{V\tablenotemark{a}} &
  \colhead{$T_{\rm eff}$, log($g$),$v_t$\tablenotemark{b}} & 
  \colhead{[Fe/H](Fe~I)} & \colhead{Date\tablenotemark{c}} & 
  \colhead{Exp. Time} & \colhead{SNR\tablenotemark{d}} & 
  \colhead{$v_r$} \\
\colhead{} & \colhead{(mag)} & \colhead{(K,~dex,~km s$^{-1}$)} &
  \colhead{(dex)} & \colhead{} & \colhead{(sec)} &
  \colhead{ } & \colhead{(km s$^{-1}$)}
}
\startdata
NGC~2419 \\ 
Stet~223\tablenotemark{e} &
         17.25 & 4250,~0.50,~2.4 & $-2.09$ & 2/2010 & 8400 & $>100$ & $-22.4$ \\
Stet~810 & 17.31 & 4330,~0.60,~2.1 & $-2.00$ & 
        11/2010 &  7200 &$>100$ & $-22.6$ \\
Stet~973\tablenotemark{f} &
      17.45 & 4320,~0.70,~2.1 & $-1.98$ & 9/2006 & 3200 & 39 & $-21.9$\\
Stet~1131 & 17.61  & 4435,~0.75,~2.2 &  $-1.95$  & 
     2/2010 & 9000 & 95 & $-16.8$  \\
Stet~1209\tablenotemark{g} &
         17.41 & 4350,~0.40,~2.1 & $-2.20$ & 11/2010 & 7200 & 93 & $-19.0$ \\
Stet~1305\tablenotemark{h} & 
      17.61 & 4400,~0.80,~2.3 & $-2.17$ & 9/2008 & 3000 & 52 & $-16.8$ \\
Stet~1814\tablenotemark{i} & 
       17.27 &  4370,~0.55,~2.3 & $-2.04$ & 2/2010 & 5400 & 90 & $-26.1$ \\
~ \\
NGC~7099 \\
Stet~31 & 13.92 & 4790,~1.63,~2.2 & $-2.39$ & 9/2006 & 1200 & $>100$ & $-185.8$ \\
Stet~34 & 13.57 & 4658,~1.13,~2.0 & $-2.32$ & 9/2006 & 1200 & $>100$ & $-188.1$ \\
AL~38\tablenotemark{j}&  13.27\tablenotemark{j} &
     4653,~1.10,~2.1  & $-2.47$ & 9/2006 & 900 & $>100$ & $-185.6$ \\
Stet~42 &  14.70 & 4935,~2.03,~1.8 & $-2.46$ & 9/2006 & 1200 & $>100$ & $-190.1$ \\
Stet~81\tablenotemark{l} & 14.25 & 5050,~1.70,~2.2 &
          $-2.48$ & 9/2006 &1200 & $>100$ & $-185.9$ \\
\enddata
\tablenotetext{a}{Star IDs and $V$~mags are from the
online version of the database of \cite{stetson05}.}
\tablenotetext{b}{These are the values adopted for the
abundance analysis.}
\tablenotetext{c}{Date of the best spectrum.  Earlier 
low SNR spectra were taken for several of the NGC~2419 stars.}
\tablenotetext{d}{SNR per spectral resolution element
in the continuum at 5500~\AA.}
\tablenotetext{e}{Stetson~223 = Suntzeff~1.}
\tablenotetext{f}{Stetson~973 = Suntzeff~15.}
\tablenotetext{g}{Stetson~1209 = Suntzeff~16.}
\tablenotetext{h}{Stetson~1305 = RH~10 
  (private communication from M.~Shetrone),
  previously observed by \cite{shetrone01}.}
\tablenotetext{i}{Stetson~1814 = Suntzeff~14.}
\tablenotetext{j}{ID  from \cite{alcaino80}.}
\tablenotetext{k}{$V$~mag from \cite{bolte87}.}
\tablenotetext{l}{S81 lies slightly bluer in $B-V$ than the mean RGB locus.}
\end{deluxetable}         


\clearpage
\begin{deluxetable}{l r rrr}
\tablenum{2}
\tablewidth{0pt}
\tablecaption{Choice of $T_{\rm eff}$ For Red Giant Members of NGC~2419
\label{table_temp} }
\tablehead{
\colhead{Name\tablenotemark{a}} & \colhead{$V$\tablenotemark{a}} &
\colhead{$T_{\rm eff}$(Adopt)} &
\colhead{$T_{\rm eff}(Phot),~\sigma(T_{\rm eff}$)\tablenotemark{b}} &
\colhead{$T_{\rm eff}$(Isochrone)} \\
\colhead{} & \colhead{(mag)} & \colhead{(K)} &
   \colhead{(K)} &   \colhead{(K)} 
}
\startdata
NGC~2419 \\
Stet~223 & 17.25 & 4250 & 4150 (20) & 4264 \\
Stet~810 & 17.31 & 4330 & 4372 (20) & 4286 \\
Stet~973 & 17.45 & 4320 & 4392 (39) & 4340\\
Stet~1131 & 17.61 & 4435 & 4428 (35)\tablenotemark{c} &  4384 \\
Stet~1209 & 17.41 & 4350 & 4328 (46) & 4327 \\
Stet~1305 &  17.61 & 4400 & 4488 (86)\tablenotemark{d} & 4385 \\
Stet~1814 &  17.27 & 4370 & 4355 (55) & 4274\\
NGC~7099 \\
S31 & 13.92 & 4790 & 4770 (24) & \nodata \\
S34 & 13.57 & 4658 & 4658 (27) & \nodata \\
AL38 & 13.27 &  4655 & 4655 (52)\tablenotemark{e} & \nodata \\
S42 &  14.70 & 4935 &  4935 (44)\tablenotemark{e} & \nodata \\
S81 & 14.25 & 5050 &  4970 (64)\tablenotemark{e} & \nodata \\
\enddata
\tablenotetext{a}{Star IDs and $V$~mags are mostly from the
online version of the
database of \cite{stetson05}.  See notes to Table~\ref{table_sample} for details.}
\tablenotetext{b}{rms of $T_{\rm eff}$ as derived from the available colors
including $V-I$, $V-J$, and $V-K_s$.}
\tablenotetext{c}{Uncertainty in 2MASS $K_s$ is 0.09~mag.}
\tablenotetext{d}{Uncertainty in 2MASS $K_s$ is 0.10~mag.}
\tablenotetext{e}{No I mag available.}
\end{deluxetable} 

\clearpage

\begin{deluxetable}{llrr rrrrr rr}     
\tabletypesize
\scriptsize              
\tablenum{3}  
\tablewidth{0pt}                                                                
\tablecaption{Equivalent Widths for NGC~2419 RGB Stars\label{table_2419_eqw}} 
\tablehead{                                                                     
\colhead{$\lambda$} &  \colhead{Ion} & \colhead{$\chi_{exc}$} & 
\colhead{log $gf$} & 
\colhead{S223} & \colhead{S1814} &  \colhead{S810} & \colhead{S1209} & \colhead{S973} 
    & \colhead{S1305}  & \colhead{S1131}  \\
\colhead{(\AA)} & \colhead{} &  \colhead{(eV)} & \colhead{(dex)} &
\colhead{(m\AA)} & \colhead{(m\AA)} & \colhead{(m\AA)} & \colhead{(m\AA)}
& \colhead{(m\AA)} & \colhead{(m\AA)} & \colhead{(m\AA)}
}
\startdata
 6707.8 & Li~I & 0.00 & 0.178 & \nodata  & \nodata  & \nodata  & \nodata  & 
      \nodata  & \nodata  & $\leq$7.0   \\
 6300.30 & [O~I] &   0.00 &  $-$9.780 &   \nodata  &   \nodata  &   \nodata  &    36.7 &    54.7 &   \nodata  &   \nodata   \\
 6363.78 & [O~I] &   0.02 & $-$10.300 &   \nodata  &   \nodata  &   \nodata  &   \nodata  &    23.5 &   \nodata  &   \nodata   \\
 5682.63 & Na~I  &   2.10 &  $-$0.700 &    30.3 &    14.9 &     6.0 &   \nodata  &   \nodata  &   \nodata  &    13.0  \\
 5688.19 & Na~I  &   2.10 &  $-$0.420 &    46.0 &   \nodata  &    15.0 &   \nodata  &   \nodata  &   \nodata  &    36.0  \\
 5889.95 & Na~I  &   0.00 &   0.110 &   \nodata  &   \nodata  &   \nodata  &   \nodata  &   362.0 &   376.0 &   \nodata   \\
 5895.92 & Na~I  &   0.00 &  $-$0.190 &   \nodata  &   \nodata  &   236.0 &   282.0 &   279.0 &   355.0 &   \nodata   \\
 4703.00 & Mg~I  &   4.34 &  $-$0.440 &    92.0 &   140.0 &    96.4 &   125.0 &   123.0 &   102.3 &    56.0  \\
\enddata
\end{deluxetable}

\clearpage

\begin{deluxetable}{llrr rrrrr}     
\tabletypesize
\scriptsize              
\tablenum{4}  
\tablewidth{0pt}                                                                
\tablecaption{Equivalent Widths For NGC~7099 RGB Stars\label{table_7099_eqw}} 
\tablehead{                                                                     
\colhead{$\lambda$} &  \colhead{Ion} & \colhead{$\chi_{exc}$} & 
\colhead{log $gf$} & 
\colhead{S31} & \colhead{S34} &  \colhead{S38} & \colhead{S42} & \colhead{S81} \\
\colhead{(\AA)} & \colhead{} &  \colhead{(eV)} & \colhead{(dex)} &
\colhead{(m\AA)} & \colhead{(m\AA)} & \colhead{(m\AA)} & \colhead{(m\AA)}
& \colhead{(m\AA)}
}
\startdata
 6300.30 & [O~I] &   0.00 &  $-$9.780 &     8.0 &     8.7 &     7.8 &     7.8 &     5.0 \\
 6363.78 & [O~I] &   0.02 & $-$10.300 &     3.4 &  \nodata  &     3.3 &  \nodata  &     6.0 \\
 7771.94 & O~I &   9.15 &   0.369 &     6.5 &     3.7 &     4.2 &     7.9 &     4.3 \\
 7774.17 & O~I &   9.15 &   0.223 &  \nodata  &  \nodata  &  \nodata  &     7.5 &  \nodata  \\
 5682.63 & Na~I  &   2.10 &  $-$0.700 &  \nodata  &     9.0 &    12.2 &  \nodata  &    10.0 \\
 5688.19 & Na~I  &   2.10 &  $-$0.420 &    11.1 &    15.0 &    21.6 &  \nodata  &    15.0 \\
 5889.95 & Na~I  &   0.00 &   0.110 &   217.1 &   236.6 &   261.9 &   171.0 &   239.0 \\
 5895.92 & Na~I  &   0.00 &  $-$0.190 &   195.9 &   210.8 &   232.7 &   157.3 &   215.2 \\
\enddata     
\end{deluxetable}

\clearpage

\begin{deluxetable}{l rrr }
\tablenum{5}
\tablewidth{0pt}
\tablecaption{Deduced Slopes for Fe~I Line-by-Line Abundances
\label{table_fe_slopes} }
\tablehead{
\colhead{ID} & \colhead{Fe~I vs EP} &
\colhead{Fe~I vs ($W_{\lambda}/{\lambda}$)} &
\colhead{Fe~I vs $\lambda$} \\
\colhead{} & \colhead{(dex eV$^{-1}$)} & \colhead{(dex)} 
    & \colhead{(dex {\AA}$^{-1}$)}
}
\startdata
NGC~2419 \\
S1131 & $-0.042$ & 0.010 & $-0.118$e$-4$ \\
S1305 & $-0.046$ & $-0.075$ &  0.168e$-4$ \\
S1209 & $-0.050$ & $-0.001$ &  0.451e$-4$ \\
S973 & $-0.038$ & 0.045 & 0.093e$-4$ \\
S1814 & $-0.083$ & $-0.001 $ & 0.169e$-4$ \\
S223 & $-0.079$ & 0.077 & $-0.136$e$-4$ \\
NGC~7099 \\
S31 & $-0.041$ & 0.008 & $-0.022$e$-4$ \\
S34 & $-0.042$ & 0.008 & $-0.173$e$-4$ \\
S38 & $-0.042$ & 0.017 & $-0.153$e$-4$ \\
S42 & $-0.055$ & 0.012 & $-0.031$e$-4$ \\
S81 & $-0.047$ & 0.009 & $-0.087$e$-5$ \\
\enddata     
\end{deluxetable}

\clearpage

\begin{deluxetable}{l rrr rrr rrr rrr rrr rrr rrr }
\rotate
\tabletypesize{\tiny}
\tablenum{6}
\tablewidth{0pt}
\tablecaption{Abundances for Red Giant Members of NGC~2419
\label{table_ngc2419_abund} }
\tablehead{
\colhead{ID\tablenotemark{a}} & \colhead{S1131} & \colhead{ } & \colhead{ } &
 \colhead{S1305} & \colhead{ } & \colhead{ } &
  \colhead{S1209} & \colhead{ } & \colhead{ } &
   \colhead{S973} & \colhead{ } & \colhead{ } &
    \colhead{S810} & \colhead{ } & \colhead{ } &
     \colhead{S1814} & \colhead{ } & \colhead{ } &
      \colhead{S223} & \colhead{ } & \colhead{} \\
\colhead{Species} & \colhead{[X/Fe]} & \colhead{$\sigma$\tablenotemark{b}} & \colhead{N\tablenotemark{c} } &
\colhead{[X/Fe]} & \colhead{$\sigma$\tablenotemark{b}} & \colhead{N} &  
\colhead{[X/Fe]} & \colhead{$\sigma$\tablenotemark{b}} & \colhead{N}  &
\colhead{[X/Fe]} & \colhead{$\sigma$\tablenotemark{b}} & \colhead{N}  &
\colhead{[X/Fe]} & \colhead{$\sigma$\tablenotemark{b}} & \colhead{N}  &
\colhead{[X/Fe]} & \colhead{$\sigma$\tablenotemark{b}} & \colhead{N}  &
\colhead{[X/Fe]} & \colhead{$\sigma$\tablenotemark{b}} & \colhead{N} \\
\colhead{ } & \colhead{(dex)} & \colhead{(dex)} & \colhead{ } &
\colhead{(dex)} & \colhead{(dex)} & \colhead{ } &
\colhead{(dex)} & \colhead{(dex)} & \colhead{ } &
\colhead{(dex)} & \colhead{(dex)} & \colhead{ } &
\colhead{(dex)} & \colhead{(dex)} & \colhead{ } &
\colhead{(dex)} & \colhead{(dex)} & \colhead{ } &
\colhead{(dex)} & \colhead{(dex)} & \colhead{ } 
}
\startdata
C\tablenotemark{d}  &  $-$0.81 & \nodata    &    1   &  $-$0.26 &   0.30  &  1   & 
    $-0.73$ & \nodata    &  1   &
   \nodata & \nodata    &  \nodata   &  $-$0.90 & \nodata    &    1   &  $-$0.79 & \nodata    &    1   & 
   \nodata & \nodata    &  \nodata   \\
O~I  & \nodata & \nodata    &  \nodata   & \nodata & \nodata    &  \nodata   & 
      0.67 & \nodata    &    1   &
     0.71 &   0.03    &    2   & \nodata & \nodata    &  \nodata   & \nodata & \nodata    &  \nodata   & 
     \nodata & \nodata    &  \nodata   \\
Na~I   &  $-$0.04 &   0.19    &    2   &   0.54 &   0.15    &    2   &   0.23 & \nodata    &    1   & 
  0.06 &   0.13    &    2   &  $-$0.42 &   0.09    &    3   &  $-$0.04 & \nodata    &    1   & 
    0.29 &   0.01    &    2   \\
Mg~I   &  $-$0.47 &   0.17    &    5\tablenotemark{e}   & 
    0.32 &   0.15    &    5   &   0.62 &   0.18    &    3   &    
  0.46 &   0.20    &    3 &   0.32 &   0.22    &    3   &   0.55 &   0.19    &    5   &
     0.30 &   0.22    &    5   \\
\enddata
\tablenotetext{a}{Star IDs from online version of \cite{stetson05}.}
\tablenotetext{b}{Dispersion about the mean abundance.}
\tablenotetext{c}{Number of detected lines.}
\tablenotetext{d}{Synthesis of CH band near 4300~\AA.}
\tablenotetext{e}{Upper limit for 5711~\AA\ Mg~I line treated as a detection.}
\tablenotetext{f}{Includes a +0.6~dex non-LTE correction for the 3961~\AA\ line. }
\tablenotetext{g}{[Fe/H] from Fe~I lines. All [X/Fe] refer to [Fe/H] from Fe~I.  The
same solar abundances as are used as in Table~\ref{table_7099_abund}.}
\end{deluxetable}

\clearpage

\begin{deluxetable}{l rrrr rrr rrr rrr rrr  }
\rotate
\tabletypesize{\scriptsize}
\tablenum{7}
\tablewidth{0pt}
\tablecaption{Abundances for Red Giant Members of NGC~7099
\label{table_7099_abund} }
\tablehead{
\colhead{ID\tablenotemark{a}} & \colhead{Stet} & \colhead{31} & \colhead{ } & \colhead{ } &
   \colhead{Stet} & \colhead{34} & \colhead{ } & 
   \colhead{AL} & \colhead{38} & \colhead{ } &
   \colhead{Stet} & \colhead{42} & \colhead{ } & 
   \colhead{Stet} & \colhead{81} & \colhead{ } \\
\colhead{Species} & \colhead{[X/Fe]} & \colhead{Log[$\epsilon$(X)]} &
     \colhead{N\tablenotemark{b}} &  \colhead{$\sigma$\tablenotemark{c}} &
  \colhead{Log[$\epsilon$(X)]} &
     \colhead{N\tablenotemark{b}} &  \colhead{$\sigma$\tablenotemark{c}} & 
   \colhead{Log[$\epsilon$(X)]} &
     \colhead{N\tablenotemark{b}} &  \colhead{$\sigma$\tablenotemark{c}} & 
   \colhead{Log[$\epsilon$(X)]} &
     \colhead{N\tablenotemark{b}} &  \colhead{$\sigma$\tablenotemark{c}} & 
   \colhead{Log[$\epsilon$(X)]} &
     \colhead{N\tablenotemark{b}} &  \colhead{$\sigma$\tablenotemark{c}} \\ 
\colhead{ } & \colhead{(dex)} & \colhead{(dex)} & \colhead{ } & \colhead{(dex)} &
    \colhead{(dex)} & \colhead{ } & \colhead{(dex)} &
    \colhead{(dex)} & \colhead{ } & \colhead{(dex)} &
    \colhead{(dex)} & \colhead{ } & \colhead{(dex)} &
    \colhead{(dex)} & \colhead{ } & \colhead{(dex)}             
}     
\startdata     
C\tablenotemark{d}  &  $-$0.11 &   6.09   &    1 &  \nodata   &   5.76   &    1 &  \nodata   &   5.48   &    1 &  \nodata   
 &   6.36   &    1 &  \nodata   &   5.73   &    1 &  \nodata   \\
O~I  &   0.76 &   7.20   &    3 &    0.08   &    6.86   &    2 &    0.21   &    6.91   &    3 &    0.17   
 &   7.26   &    3 &    0.13   &   6.65   &    1 &  \nodata   \\
Na~I   &   0.20 &   4.13   &    4 &    0.10   &    4.18   &    4 &    0.09   &     4.33   &    4 &    0.14   
 &   3.89   &    2 &    0.11    &   4.49   &    4 &    0.22   \\
Mg~I   &   0.53 &   5.68   &    7 &    0.17   &    5.82   &    7 &    0.12   &     5.53   &    7 &    0.08 
 &   5.57   &    7 &    0.20    &   5.41   &    7 &    0.17   \\
Al~I\tablenotemark{e}   &   0.21 &   4.29   &    1 &  \nodata   &     4.75   &    1 &  \nodata   &      4.82   &    1 &  \nodata  
 &   4.07   &    1 &  \nodata   &     4.85   &    2 &    0.11   \\
Si~I   &   0.46 &   5.62   &    8 &    0.11   &     5.69   &    9 &    0.10   &     5.64   &    9 &    0.10  
 &   5.62   &    4 &    0.18   &    5.69   &    6 &    0.12   \\
K~I    &   0.57 &   3.30   &    1 &  \nodata   &    3.48   &    1 &  \nodata   &   3.19   &    1 &  \nodata   
 &   3.15   &    1 &  \nodata    &   3.39   &    1 &  \nodata   \\
Ca~I   &   0.28 &   4.25   &   21 &    0.16   &   4.31   &   23 &    0.15   &     4.15   &   21 &    0.15   
 &   4.24   &   20 &    0.18    &   4.20   &   20 &    0.14   \\
Sc~II  &   0.23 &   0.94   &   10 &    0.10   &   0.91   &   10 &    0.11   &     0.88   &   10 &    0.08 
 &   0.89   &    7 &    0.14   &    0.87   &    8 &    0.13   \\
Ti~I   &   0.21 &   2.81   &   22 &    0.10   &     2.84   &   18 &    0.06   &    2.68   &   18 &    0.09   
 &   2.76   &   17 &    0.09   &   2.79   &   17 &    0.05   \\
Ti~II  &   0.41 &   3.01   &   22 &    0.13   &      2.97   &   20 &    0.14   &    2.97   &   20 &    0.15   
 &   2.99   &   21 &    0.13   &     2.87   &   24 &    0.15   \\
V~I    &  $-$0.22 &   1.39   &    1 &  \nodata   &  1.59   &    3 &    0.06   &     1.20   &    1 &  \nodata   
 &   1.27   &    1 &  \nodata   &    1.28   &    1 &  \nodata   \\
Cr~I   &  $-$0.32 &   2.96   &    9 &    0.09   &     3.07   &   10 &    0.11   &    2.88   &   10 &    0.13   
 &   2.95   &   10 &    0.08   &    2.88   &   10 &    0.12   \\
Mn~I   &  $-$0.39 &   2.61   &    6 &    0.06    &   2.57   &    7 &    0.05   &    2.59   &    7 &    0.08    
 &   2.56   &    6 &    0.13   &    2.50   &    6 &    0.13   \\
Fe~I\tablenotemark{f}   & $-2.39$ &   5.06   &  121 & 0.13   &
       5.13   &  125 &    0.14    &   4.97   &  115 &    0.14   &  
     4.99   &   99 &    0.15   &   4.97   &  107 &    0.13   \\
Fe~II  &   0.05 &   5.11   &   16 &    0.14   &   5.11   &   20 &    0.11   &   5.09   &   20 &    0.13  
 &   5.05   &   15 &    0.19   &    5.01   &   16 &    0.14   \\
Co~I   &  $-$0.26 &   2.27   &    1 &  \nodata   &    2.29   &    2 &    0.05    &   2.33   &    2 &    0.16   
 &   2.39   &    1 &  \nodata   &   2.42   &    1 &  \nodata   \\
Ni~I   &   0.02 &   3.88   &   14 &    0.18    &   3.87   &   16 &    0.18   &   3.77   &   16 &    0.16 
 &   3.86   &    9 &    0.23   &   3.86   &    6 &    0.19   \\
Cu~I   &  $-$0.77 &   1.05   &    1 &  \nodata   &   1.12   &    1 &  \nodata   &    1.01   &    1 &  \nodata  
 & \nodata   &  \nodata &  \nodata   & \nodata & \nodata     \\
Zn~I   &   0.14 &   2.35   &    2 &    0.11    &   2.33   &    2 &    0.05    &   2.29   &    2 &    0.01 
 &   2.33   &    2 &    0.05    &   2.37   &    2 &    0.17   \\
Sr~II  &   0.01 &   0.52   &    2 &    0.14   &   0.45   &    2 &    0.16  &   0.52   &    2 &    0.06 
 &   0.52   &    2 &    0.12  &   0.48   &    2 &    0.05   \\
Y~II   &  $-$0.10 &  $-$0.25   &    6 &    0.08  &  $-$0.41   &    6 &    0.02   &   $-$0.50   &    6 &    0.15  
 &  $-$0.26   &    6 &    0.15   &  $-$0.32   &    5 &    0.12   \\
Zr~II  &   0.33 &   0.54   &    3 &    0.13   &   0.49   &    3 &    0.15   &   0.44   &    3 &    0.16  
 &   0.66   &    3 &    0.18  &   0.47   &    3 &    0.08   \\
Ba~II  &  $-$0.10 &  $-$0.35   &    4 &    0.08   &  $-$0.44   &    4 &    0.07   &  $-$0.41   &    4 &    0.07   
 &  $-$0.44   &    4 &    0.12    &  $-$0.45   &    4 &    0.06   \\
La~II  &  $-$0.15 &  $-$1.40   &    4 &    0.22   &    $-$1.26   &    5 &    0.08  &    $-$1.37   &    5 &    0.05  
 &  $-$1.07   &    4 &    0.13   &  $-$1.20   &    4 &    0.13   \\
Ce~II  &   0.11 &  $-$0.73   &    4 &    0.23   &  $-$0.92   &    5 &    0.16   &   $-$1.04   &    4 &    0.11  
 &  $-$0.71   &    1 &  \nodata    &  $-$0.45   &    1 &  \nodata   \\
Nd~II  &   0.25 &  $-$0.64   &    7 &    0.14   &    $-$0.81   &   10 &    0.10   &  $-$0.77   &   10 &    0.08  
 &  $-$0.63   &    4 &    0.18   &   $-$0.83   &    2 &    0.09   \\
Sm~II  & \nodata & \nodata   &  \nodata &  \nodata   & \nodata & \nodata   &  \nodata &   $-$1.30   &    2 &    0.02  
 & \nodata   &  \nodata &  \nodata   & \nodata & \nodata   &  \nodata    \\
Eu~II  &   0.40 &  $-$1.48   &    2 &    0.04   &   $-$1.72   &    4 &    0.20   &    $-$1.45   &    3 &    0.05  
 &  $-$1.38   &    2 &    0.04    &  $-$1.51   &    2 &    0.01   \\
Dy~II  &   0.64 &  $-$0.65   &    1 &  \nodata   &  $-$0.68   &    2 &    0.27   &  $-$0.98   &    2 &    0.30 
 &  $-$0.71   &    1 &  \nodata   &  $-$1.13   &    1 &  \nodata   \\
\enddata
\tablenotetext{a}{Star IDs from online version of \cite{stetson05} or from 
   \cite{alcaino80}.} 
\tablenotetext{b}{Number of detected lines.}
\tablenotetext{c}{Dispersion about the mean abundance.}
\tablenotetext{d}{Synthesis of CH band near 4300~\AA.}
\tablenotetext{e}{Includes a +0.6~dex non-LTE correction for the 3961~\AA\ line.
Including this correction, the very marginal detections of the 6696~\AA\ line give
consistent Al abundances.}
\tablenotetext{f}{[Fe/H] from Fe~I lines. All [X/Fe] refer to [Fe/H] from Fe~I.}
\end{deluxetable}

\clearpage

\begin{deluxetable}{l rrrr}
\tablenum{8a}
\tablewidth{0pt}
\tablecaption{Abundance Sensitivities -- Absolute Abundances [X/H]
\label{table_abund_sens_abs} }
\tablehead{
\colhead{Species} & \colhead{$T_{\rm eff}$} & \colhead{log($g$)} &
     \colhead{[Fe/H](model)} & \colhead{$v_t$} \\
\colhead{} & \colhead{(+250~K)\tablenotemark{a}}  & 
   \colhead{($-$0.5~dex)\tablenotemark{b}} &
   \colhead{($-$0.5~dex)} & \colhead{(+0.2~km s$^{-1}$)} \\
\colhead{} & \colhead{(dex)} & \colhead{(dex)}  & \colhead{(dex)} & \colhead{(dex)}
}
\startdata
C(CH)\tablenotemark{c} & 0.28 & 0.15 & $-0.13$ & $-0.07$ \\
O~I\tablenotemark{d}   &  0.04 &  $-$0.21 &  $-$0.13 &  $-$0.01  \\
Na~I  & 0.57 &   0.22 &   0.06 &  $-$0.10  \\
Mg~I  &  0.23 &   0.16 &   0.18 &  $-$0.08  \\
Al~I  &  0.34 &   0.20 &   0.14 &  $-$0.04  \\
K~I   &  0.42 &   0.05 &   0.13 &  $-$0.07  \\
Ca~I  &  0.28 &   0.09 &   0.12 &  $-$0.04  \\
Sc~II &  0.00 &  $-$0.16 &  $-$0.08 &  $-$0.03  \\
Ti~I  &  0.63 &   0.08 &   0.21 &  $-$0.05  \\
Ti~II & $-$0.03 &  $-$0.04 &   0.01 &  $-$0.10  \\
V~I   &  0.54 &   0.06 &   0.14 &   0.00  \\
Cr~I  &  0.55 &   0.12 &   0.23 &  $-$0.08  \\
Mn~I  &  0.42 &   0.11 &   0.17 &  $-$0.04  \\
Fe~I  &  0.42 &   0.06 &   0.13 &  $-$0.07  \\
Fe~II & $-$0.17 &  $-$0.08 &   0.00 &  $-$0.09  \\
Co~I  &  0.45 &   0.05 &   0.11 &  $-$0.01  \\
Ni~I  &  0.42 &   0.02 &   0.09 &  $-$0.02  \\
Cu~I  &  0.44 &   0.06 &   0.12 &  $-$0.01  \\
Sr~I  &  0.44 &   0.12 &   0.17 &   0.00  \\
Sr~II &  0.10 &   0.04 &  $-$0.02 &  $-$0.07  \\
Y~II  &  0.02 &  $-$0.11 &  $-$0.03 &  $-$0.03  \\
Ba~II & 0.13 &  $-$0.16 &  $-$0.08 &  $-$0.14  \\
La~II &  0.09 &  $-$0.14 &  $-$0.06 &  $-$0.02  \\
Ce~II &  0.07 &  $-$0.10 &  $-$0.03 &  $-$0.02  \\
Nd~II &  0.10 &  $-$0.14 &  $-$0.07 &  $-$0.02  \\
Eu~II &  0.06 &  $-$0.08 &  $-$0.02 &  $-$0.09  \\
\enddata
\tablenotetext{a}{$T_{\rm eff}$ changes from 4250 to 4500~K.}
\tablenotetext{b}{{L}og($g$) changes from 1.0 to 0.5~dex.}
\tablenotetext{c}{The 4320~\AA\ region of the G band of CH was used.}
\tablenotetext{d}{Values for the [O~I] lines 6300, 6363~\AA.}
\end{deluxetable} 

\clearpage

\begin{deluxetable}{l r | rrr rr}
\tablenum{8b}
\tablewidth{0pt}
\tablecaption{Abundance Ratio Sensitivities 
\label{table_abund_sens_rel} }
\tablehead{
\colhead{Species} & \colhead{rms Total\tablenotemark{a}}  &
     \colhead{${\Delta}T_{\rm eff}$} & \colhead{$\Delta$log($g$)} &
     \colhead{${\Delta}$[Fe/H](model)} & \colhead{${\Delta}v_t$} &
     \colhead{$W_{\lambda}$\tablenotemark{b} } \\
\colhead{} &  \colhead{} & \colhead{(100~K)}  & 
   \colhead{(0.1~dex)\tablenotemark{b} } &
   \colhead{(0.1~dex)} & \colhead{(+0.2~km s$^{-1}$)} &
   \colhead{} \\
\colhead{} & \colhead{(dex)} & \colhead{(dex)}  & \colhead{(dex)} &
     \colhead{(dex)} & \colhead{(dex)} & \colhead{(dex)}
}
\startdata
C(CH)\tablenotemark{c} & 0.11 & $-0.06$ & 0.02 & $-0.05$ & 0.00 & 0.08 \\
$[$O~I/Fe~I$]$\tablenotemark{d}  &   0.13  &   0.08  &   $-$0.03   & $-$0.05 &   $-$0.01  &   0.08 \\ 
$[$Na~I/Fe~I$]$ & 0.15  &   0.06  &   0.03 &   $-$0.03  &  $-$0.10 &    0.08     \\
$[$Mg~I/Fe~I$]$ &    0.12  &  $-$0.08  &   0.02  &   0.02  &  $-$0.08 &    0.05 \\  
$[$Al~I/Fe~I$]$ &   0.08 &   $-$0.03 &   0.03  &   0.00 &   $-$0.04 & 0.06 \\   
$[$K~I/Fe~I$]$ &    0.11  &   0.00  &  0.00   &  0.00  &  $-$0.07 &    0.08 \\    
$[$Ca~I/Fe~I$]$ &    0.07  &   $-$0.06   &   0.01  &   0.00    & $-$0.04  &    0.02 \\
$[$Sc~II/Fe~II$]$ &  0.09   &  0.07   & $-$0.02 &   $-$0.03  &  $-$0.03   &  0.03 \\  
$[$Ti~I/Fe~I$]$  &   0.11   &  0.08  &   0.00 &    0.03 &   $-$0.05  &   0.02 \\ 
$[$Ti~II/Fe~II$]$  &    0.12   &   0.06  &    0.01   &   0.00 &   $-$0.10   &   0.03 \\
$[$V~I/Fe~I$]$ &   0.07  &   0.05 &    0.00 &    0.00  &   0.00  &   0.05  \\ 
$[$Cr~I/Fe~I$]$ &    0.11  &    0.05 &     0.01 &     0.04 &    $-$0.08   &   0.03 \\  
$[$Mn~I/Fe~I$]$ &   0.06  &    0.00  &    0.01 &     0.02 &    $-$0.04  &    0.03 \\  
$[$Fe~I/Fe~I$]$ &    0.07 &    0.00  &   0.00  &   0.00  &  $-$0.07  &   0.01 \\ 
$[$Fe~II/Fe~I$]$  &  0.26  &  $-$0.24  &  $-$0.03  &  $-$0.05  &  $-$0.09   &  0.03  \\ 
$[$Fe~II/Fe~II$]$ &    0.09 &    0.00  &   0.00   &  0.00  &  $-$0.09 &   0.03  \\
$[$Co~I/Fe~I$]$ &  0.08  &   0.01 &  0.00 &   $-$0.01  &  $-$0.01   &  0.08 \\ 
$[$Ni~I/Fe~I$]$ &   0.04  &     0.00   &   $-$0.01   &   $-$0.02   &   $-$0.02  &     0.02 \\
$[$Cu~I/Fe~I$]$ & 0.08  &   0.01  &   0.00  &  0.00 &   $-$0.01  &   0.08 \\  
$[$Sr~I/Fe~I$]$ &  0.08  &    0.01   &   0.01 &     0.02 &     0.00 &     0.08 \\   
$[$Sr~II/Fe~II$]$ &   0.15  &    0.11 &    0.02 &   $-$0.01 &    $-$0.07 &     0.08 \\   
$[$Y~II/Fe~II$]$ &   0.09 &   0.08 &    $-$0.01 &    $-$0.01 &   $-$0.03 &    0.04 \\  
$[$Ba~II/Fe~II$]$ &   0.19 &    0.12 &   $-$0.02 &    $-$0.03 &   $-$0.14 &   0.05 \\  
$[$La~II/Fe~II$]$ &   0.14 &    0.10 &   $-$0.01 &   $-$0.02 &   $-$0.02 &    0.08 \\ 
$[$Ce~II/Fe~II$]$ &  0.11 &   0.10 &  0.00 &    $-$0.01 &  $-$0.02 &    0.06 \\ 
$[$Nd~II/Fe~II$]$ & 0.13 &   0.11 &   $-$0.01 &   $-$0.03 &   $-$0.02 &    0.06 \\ 
$[$Eu~II/Fe~II$]$ &  0.14 &   0.09 &    0.00  &  $-$0.01 &    $-$0.09 &   0.06 \\  
\enddata
\tablenotetext{a}{The rms sun of all 5 contributing terms, relative
to whichever of Fe~I or Fe~II gave the smaller value.}
\tablenotetext{b}{Assumed to be 0.08~dex/$\sqrt{N(lines)}$.}
\tablenotetext{c}{The 4320~\AA\ region of the G band of CH was used.}
\tablenotetext{d}{Values for the [O~I] lines 6300, 6363~\AA.}
\end{deluxetable}

\clearpage

\begin{deluxetable}{l rrr | rrr}
\tablenum{9}
\tablewidth{0pt}
\tablecaption{Mean Abundances for Red Giant Members of NGC~2419
  and of NGC~7099
\label{table_abund_mean} }
\tablehead{
\colhead{} & \colhead{NGC} & \colhead{7099} & \colhead{} & 
   \colhead{NGC} & \colhead{2419} \\
\colhead{Species} &
\colhead{$<$[X/Fe]$>$\tablenotemark{a}} & 
    \colhead{$\sigma$[X/Fe]} & \colhead{N. Stars} & 
  \colhead{$<$[X/Fe]$>$\tablenotemark{a}} & 
    \colhead{$\sigma$[X/Fe]} &  \colhead{N. Stars} \\
\colhead{} & \colhead{(dex)} & \colhead{(dex)} & \colhead{} &
   \colhead{(dex)} & \colhead{(dex)} & \colhead{}
}
\startdata
log[$\epsilon$(Fe)](FeI) & $-2.42$ & 0.07 & 5 & $-2.06$ & 0.10 & 7 \\
log[$\epsilon$(Fe)](FeII) & $-2.38$ & 0.04 & 5 & $-2.12$ & 0.07 & 7 \\
$[$C/Fe$]$ & $-0.28$ & 0.35 & 5 & $-0.70$ & 0.25 & 5 \\
$[$O/Fe$]$ & 0.35 & 0.28 & 5 & 0.74 & 0.10 & 2 \\
$[$Na/Fe$]$ & 0.29 & 0.27 & 5 & 0.09 & 0.31 & 7 \\
$[$Mg/Fe$]$ & 0.51 & 0.11 & 5 & 0.30 & 0.36 & 7 \\
$[$Al/Fe$]$\tablenotemark{b} & 0.51 & 0.36 & 5 
      & 0.45 & \nodata & 1 \\
$[$Si/Fe$]$ & 0.52 & 0.07 & 5 & 0.43 & 0.12 & 5  \\
$[$K/Fe$]$ & 0.61 & 0.10 & 5 & 0.58 & 0.28 & 6 \\
$[$Ca/Fe$]$ & 0.29 & 0.05 & 5 & 0.15 & 0.07 & 7 \\
$[$ScII/FeII$]$ & 0.18 & 0.03 & 5 & 0.17 & 0.14 & 7 \\
$[$Ti/Fe$]$ & 0.21 & 0.05 & 5 & 0.11 & 0.07 & 7 \\
$[$TiII/FeII$]$ & 0.35 & 0.03  & 5 & 0.30 & 0.16 & 7 \\
$[$V/Fe$]$ & $-0.23$ & 0.09 & 5 & 0.03 & 0.09 & 7 \\
$[$Cr/Fe$]$ & $-0.30$ & 0.03 & 5 & $-0.23$ & 0.09 & 7 \\
$[$Mn/Fe$]$ & $-0.40$ & 0.06 & 5 & $-0.28$ & 0.07 & 7 \\
$[$FeII/FeI$]$ & 0.05 & 0.05 & 5 & $-0.06$ & 0.05 & 7 \\
$[$Co/Fe$]$ & $-0.16$ & 0.12 & 5 & 0.09 & 0.08 & 4 \\
$[$Ni/Fe$]$ & 0.02 & 0.07 & 5 & $-0.05$ & 0.03 & 7 \\
$[$Cu/Fe$]$ & $-0.76$ & 0.17 & 3 & $-0.61$ & 0.10 & 7 \\
$[$Zn/Fe$]$ & 0.16 & 0.07 & 5 & $-0.20$ & 0.06 & 4 \\
$[$Sr/Fe$]$ & \nodata & \nodata & \nodata &
           $-0.23$ & 0.13 & 4 \\
$[$SrII/FeII$]$ & $-0.03$ & 0.05 & 5 & $-0.08$ & 0.12 & 3 \\
$[$YII/FeII$]$ & $-0.23$ & 0.15 & 5 & $-0.36$ & 0.09 & 7 \\
$[$BaII/FeII$]$ & $-0.18$ & 0.04 & 5 & $-0.11$ & 0.14 & 7 \\
$[$LaII/FeII$]$ & $-0.02$ & 0.17 & 5 & 0.05 & 0.30 & 4 \\
$[$NdII/FeII$]$ & 0.14 & 0.10 & 5 & 0.00 & 0.23 & 6 \\
$[$EuII/FeII$]$ & 0.41 & 0.18 & 5 & 0.30 & 0.15 & 5 \\
\enddata
\tablenotetext{a}{$[$X/Fe$]$ where both species are neutral, unless
otherwise indicated.}
\tablenotetext{b}{A non-LTE correction of +0.6~dex 
has been applied to abundances derived from the 3961~\AA\ line of Al~I.}
\end{deluxetable} 

\clearpage

\begin{deluxetable}{l rr c}
\tablenum{10}
\tablewidth{0pt}
\tablecaption{Comparison of Our Mean Abundances for Red Giant Members of NGC~7099
   With Those From the Carretta et al VLT Large Program
\label{table_7099_comp} }
\tablehead{
\colhead{Species} & \colhead{Carretta et al\tablenotemark{ab}} & 
   \colhead{This Paper\tablenotemark{c}} & \colhead{Notes}  \\
\colhead{} & \colhead{(dex)} & \colhead{(dex)}  & \colhead{} \\
\colhead{} & \colhead{(Mean,$\sigma$)} & \colhead{(Mean,$\sigma$)} & \colhead{}
}
\startdata
$[$Fe/H$]$  &  $-2.35$ (0.05) & $-2.42$ (0.07) & \\
$[$O/Fe$]$ &  0.46 (0.20) & 0.35 (0.28) & d, e \\
$[$Na/Fe$]$ & 0.35 (0.25) & 0.29 (0.27) & d \\
$[$Mg/Fe$]$ & 0.51 (0.04) & 0.51 (0.11) \\
$[$Al/Fe$]$ & 0.77 (0.32) & 0.51 (0.36) & d, e \\
$[$Si/Fe$]$ & 0.34 (0.07) & 0.52 (0.07) \\
$[$Ca/Fe$]$ & 0.31 (0.03)  & 0.29 (0.05) \\
$[$Ba/Fe$]$ &  $-0.18$ (0.04) & $-0.22$ (0.14) & f, g \\
\enddata
\tablenotetext{a}{Carretta et al 2009abc, Carretta et al 2010.}
\tablenotetext{b}{UVES sample of 10 RGB stars in NGC~7099.}
\tablenotetext{c}{Our sample of 5 RGB stars.}
\tablenotetext{d}{This element has a large internal abundance range.}
\tablenotetext{e}{There are many upper limits in the VLT 
sample, which E.~Carretta (private communication)
advises they treated as  detections
to calculate their mean abundance ratios.}
\tablenotetext{f}{From the large intermediate resolution FLAMES/VLT
survey, see \cite{gc_bastars}.}
\tablenotetext{g}{This, for our result, is the ratio $[$Ba~II/Fe~II$]$.}
\end{deluxetable} 

\clearpage

\end{document}